\RequirePackage[l2tabu, orthodox]{nag}
\documentclass{article}
\usepackage[english]{babel}
\usepackage[top=2cm,bottom=2cm,left=2cm,right=2cm]{geometry}
\usepackage{amsmath,amsthm,amssymb,amsfonts}
\usepackage{graphicx} 
\usepackage{epstopdf}
\usepackage{syntonly}
\usepackage{subfiles}
\usepackage{caption}
\usepackage{subcaption}
\usepackage{booktabs}
\usepackage{listings}
\usepackage{xcolor}
\usepackage{tabularray}
\usepackage{multirow}
\UseTblrLibrary{siunitx}
\usepackage{algorithmic}
\usepackage{algorithm}
\usepackage{array}
\usepackage{float}
\usepackage{hyperref} 
\usepackage{cleveref}
\usepackage[sort]{natbib}
\setcitestyle{numbers,square,sort}
\usepackage{authblk}

\newtheorem{definition}{Definition}
\newcommand{\refeq}[1]{Eq.(\ref{#1})}
\newcommand{\reffig}[1]{Figure \ref{#1}}
\newcommand{\reftab}[1]{Table \ref{#1}}

\title{\textbf{Mitigating Extremal Risks: A Network-Based Portfolio Strategy}}

\author[a]{Qian Hui \thanks{\href{mailto:qhui24@m.fudan.edu.cn}{qhui24@m.fudan.edu.cn}.}}
\author[a,b]{Tiandong Wang \thanks{Corresponding author, \href{mailto:td_wang@fudan.edu.cn}{td\underline{ }wang@fudan.edu.cn}.}}

\affil[a]{Shanghai Center for Mathematical Sciences, Fudan University}
\affil[b]{Shanghai Academy of Artificial Intelligence for Science}

\date{}
\begin{document}
\maketitle

\begin{abstract}
In financial markets marked by inherent volatility, extreme events can result in substantial investor losses. This paper proposes a portfolio strategy designed to mitigate extremal risks. By applying extreme value theory, we evaluate the extremal dependence between stocks and develop a network model reflecting these dependencies. We use a threshold-based approach to construct this complex network and analyze its structural properties. To improve risk diversification, we utilize the concept of the maximum independent set from graph theory to develop suitable portfolio strategies. Since finding the maximum independent set in a given graph is NP-hard, we further partition the network using either sector-based or community-based approaches. Additionally, we use value at risk and expected shortfall as specific risk measures and compare the performance of the proposed portfolios with that of the market portfolio.

\end{abstract}

\section{Introduction}
Financial markets are inherently volatile, leading to sudden and extreme fluctuations that can severely impact investor portfolios. These extreme events, such as market crashes or sharp downturns, present serious challenges to conventional risk management strategies, often resulting in substantial financial losses. When modeling extremal risks, traditional correlation measures often fail in the presence of extremal dependence, as the second moment of a heavy-tailed random variable may not exist. 
Further limitations of correlation-based methods have been discussed in \cite{poon2004extreme, kelly2014tail}.

Therefore, various measures of extremal dependence have been studied in the literature. 
Among existing studies, two popular ways to quantify the extremal dependence are the \emph{extremal dependence measure} (EDM) \cite{resnick2004extremal} and \emph{extremograms} \cite{davis2009extremogram}. 
The EDM is a statistical tool that quantifies the tendency of large values of components of a random vector to occur simultaneously, and the extremogram describes how extreme events (such as large losses or gains) at one time point relate to extreme events at another time point.
A detailed comparison between these two tools is provided in \cite{EDM}. 
With these analytical tools available, the key question becomes: How can they guide investors in constructing portfolios that are resilient to extremal losses in the market?
Hence, in the current study, our primary aim is to first quantify the degree of pairwise extremal dependence to generate a dependence network of stock returns, and then apply graphical tools to find the optimal strategy for hedging against extremal risk. 
We utilize the extremal dependence measure \cite{resnick2004extremal} to characterize the degree of dependence between pairs of heavy-tailed random variables.




In the constructed extremal dependence network, as long as there is no edge between two nodes, we regard these two stock returns as having low extremal dependence. 
Then we integrate the concept of the maximum independent set (MIS) from graph theory into our portfolio optimization process. 
The MIS represents a group of stock returns with minimum extremal dependence, making them ideal for constructing diversified portfolios that are more resilient to extremal risks. 
In \cite{boginski2005statistical} and \cite{spelta2022chaos}, the authors showcase the importance of using MIS in financial networks to enhance portfolio robustness. Since finding the MIS is NP-hard, we further partition the network into smaller sub-networks based on economic sectors or community structures, facilitating the identification of local MIS and enabling efficient portfolio construction.

The rest of this paper is organized as follows. Section \ref{sec2} outlines multivariate regular variation, the EDM, and its application to stock price fluctuation dependence. Section \ref{sec3} details the construction of a stock network model based on extremal dependence. Section \ref{sec4} presents empirical findings and a comparison of local and overall portfolios, demonstrating the strategies' effectiveness in mitigating extremal risks for investors. The paper concludes in Section \ref{sec5}.

\subsection{Data example}
We use the R package \verb6quantmod6 to retrieve data from Yahoo Finance on 113 constituent stocks from the Shenzhen component of the CSI 300 in 2023. The trading period spans from January 1, 2023 to December 31, 2023, with a total of 242 trading days.
We compute the log-return of stock $i$ on day $t$ as
\[
	r_i(t):=\log P_i(t) -\log P_{i}(t-1),
\]
where $P_{i}(t)$ represents the adjusted closing price of stock $i$ on day $t$. Next, we calculate the extremal dependence measure (EDM) between stock $p$ and stock $q$,  $\text{EDM}(p,q)$, by substituting the returns into \refeq{eq2.5}. 

Combined with the data above, we propose an algorithm below to construct stock portfolios based on their extremal dependence structures and then give the optimal portfolio.
\begin{algorithm}[H]
    \caption{Portfolio construction using EDM.}
    \renewcommand{\algorithmicrequire}{\textbf{Input:}}
    \renewcommand{\algorithmicensure}{\textbf{Output:}}
    \begin{algorithmic}
        \REQUIRE Adjusted price of each stock $P_i(t)$, $i=1,\ldots,n$, at time $t$.
        
        Step 1: Compute the log return $r_i(t)$ for stock $i$, and then calculate the pairwise EDM based on \refeq{eq2.5};
        
        Step 2: Denote each stock as a vertex, and use a threshold-based approach to construct networks;
        
        Step 3: Divide networks into clusters using a proper criterion such as by sector or by community;
        
        Step 4: Solve for the maximum independent set of each cluster;

        Step 5: Use risk measurement indicators such as VaR or ES for each maximum independent set, and construct a portfolio optimization model by minimizing the overall risk.
        
        \ENSURE Optimal portfolio with minimum risk.
    \end{algorithmic}
\end{algorithm}

\section{Extremal dependence between stock returns}\label{sec2}
The extremal dependence measure (EDM) (cf. \cite{EDM}) quantifies the tendency for large values to occur simultaneously between two components, and we further use it as our main tool to construct the network structure between stock returns.

We start by introducing the definition of regular variation.
In one dimension, a measurable function \(f\) is regularly varying with index $\alpha$, \(\alpha \in \mathbb{R} \)  if $f: \, \mathbb{R}_{+} \mapsto \mathbb{R}_{+}$ satisfies
\begin{equation}
	\lim_{t\rightarrow \infty} \frac{f(tx)}{f(t)} = x^{\alpha}, \,\, \text{for} \,\, x>0,
\end{equation}
denoted as $f \in RV_{\alpha}$. To formalize our analysis, we provide some useful definitions related to multivariate regular variation (MRV) of measures, and it is a natural extension of the one-dimensional regular variation. 

Suppose that $\mathbb{C}_0 \subset \mathbb{C} \subset \mathbb{R}_{+}^2$ are two closed cones, and we provide the definition of $\mathbb{M}$-convergence in Definition \ref{def1} (cf. \cite{basrak2019note, das2013living, hult2006regular, kulik2020heavy, Lindskog2013RegularlyVM}) on $\mathbb{C} \setminus \mathbb{C}_0$, which lays the theoretical foundation of regularly varying measures (cf. Definition \ref{def2}).

\begin{definition}\label{def1}
Let $\mathbb{M}(\mathbb{C} \setminus \mathbb{C}_0)$ be the set of Borel measures on $\mathbb{C} \setminus \mathbb{C}_0$ which are finite on sets bounded away from $\mathbb{C}_0$, and $\mathbb{C} \setminus \mathbb{C}_0$ be the set of continuous, bounded, non-negative functions on $\mathbb{C} \setminus \mathbb{C}_0$ whose supports are bounded away from $\mathbb{C}_0$. Then for $\mu_n$, $\mu \in \mathbb{M}(\mathbb{C} \setminus \mathbb{C}_0)$, we say $\mu _n \rightarrow \mu $ in $\mathbb{M}(\mathbb{C} \setminus \mathbb{C}_0)$, if $\int f\text{d}\mu_n \rightarrow \int f\text{d}\mu$ for all $f \in \mathcal{C} (\mathbb{C} \setminus \mathbb{C}_0)$.
\end{definition}

\begin{definition}\label{def2}
The distribution of a random vector $\textbf{Z}=[Z_1,Z_2]^T$ on $\mathbb{R}_{+}^2$, i.e. $\mathbb{P}(\textbf{Z} \in \cdot)$, is (standard) regularly varying on $\mathbb{C} \setminus \mathbb{C}_0$ with index $c>0$ (written as $\mathbb{P}(\textbf{Z} \in \cdot) \in \text{MRV}(c,b(t),\nu,\mathbb{C} \setminus \mathbb{C}_0)$) if there exists some scaling function $b(t) \in RV_{1/c}$ and a limit measure $\nu(\cdot) \in \mathbb{M}(\mathbb{C} \setminus \mathbb{C}_0)$ such that as $t\rightarrow \infty$,
\begin{equation}\label{eq2.1}
    t\mathbb{P}\left( \frac{\textbf{Z}}{b(t)} \in \cdot\right)  \rightarrow \nu(\cdot), \,\, \text{in} \,\, \mathbb{M}(\mathbb{C} \setminus \mathbb{C}_0).
\end{equation}
\end{definition}
In \refeq{eq2.1}, all elements are normalized by the same function column $b(t)$, which implies that all marginal distributions are tail-equivalent with index $-\alpha$ \cite{EDMbook}. When analyzing the asymptotic dependence between components of a bivariate random vector $\textbf{Z}$ satisfying \refeq{eq2.1}, it is often informative to make a polar coordinate transform and consider the transformed points located on the $L_2$ unit sphere
\begin{equation}
    (x,y) \mapsto \left( \frac{x}{\sqrt{x^2+y^2}}, \frac{y}{\sqrt{x^2+y^2}} \right) ,
\end{equation}
after thresholding the data according to the $L_2$ norm. In $\mathbb{R}_{+}^2$, the convenient version of the $L_2$-polar coordinate transformation is $T:\, \textbf{Z} \mapsto (\lVert \textbf{Z} \lVert , \textbf{Z} / \lVert \textbf{Z} \lVert ) = (R, \Theta)$, we provide the following equivalent definition in polar coordinates. 

\begin{definition}(cf. \cite[Theorem 6.1]{EDMbook}) 
A 2-dimensional random vector $\textbf{Z}=[Z_1,Z_2]^T$ is (standard) regularly varying if and only if there exists a function sequence $b(t) \rightarrow \infty$ and a spectral measure $\Gamma$ on $\aleph _{+}^2 = \{\textbf{x} \in \mathbb{R}_{+}^2 \setminus \{0\} :\, \lVert \textbf{x} \lVert = 1\}$, and there exists a constant $c = \nu \{\textbf{x}: \lVert \textbf{x} \lVert > 1\} > 0$ such that
	\begin{equation}\label{eq2.2}
		t\mathbb{P}\left( \left( \frac{R}{b(t)}, \Theta \right)  \in \cdot\right)  \rightarrow c\nu_{\alpha} \times \Gamma, \,\, \text{in} \,\, \mathbb{M}_{+}(\left( 0, \infty \right] \times \aleph_{+}^2),
	\end{equation}
where $\nu_{\alpha} \left( x, \infty \right] = x^{-\alpha}$, $x>0$.
\end{definition}


Now we focus on the extremal dependence measure. Given a regularly varying bivariate random vector $\textbf{Z} = [Z_1,Z_2]^T$, the EDM is defined as (cf. \cite{EDM}, Eq.(8))
\begin{equation}\label{eq2.3}
	\text{EDM}(Z_1,Z_2) = \int _{\aleph^2_{+}} a_1a_2 \Gamma (d\textbf{a}).
\end{equation}
Notice that the minimum value of EDM is 0 if and only if the coordinates of $\textbf{Z}$ are asymptotically independent, i.e., the spectral measure $\Gamma$ concentrates on $\{ (1,0) / \lVert (1,0) \lVert , (0,1) / \lVert (0,1) \lVert \}$, or equivalently, the limit measure $\nu$ concentrates on the axes. In addition, if the norm is symmetric, then EDM reaches its maximum value if and only if the support of $\Gamma$ is $\{\textbf{a}:\, a_1=a_2 \}$, or equivalently, $\nu$ concentrates on the line $\{t(1,1),t>0\}$. 

In \cite{EDM}, the authors highlight that $\text{EDM}$ can be interpreted as the limit of the cross moment between normalized $Z_1$ and $Z_2$ when $R = \lVert \textbf{Z} \lVert $ is large, i.e.
\begin{equation}\label{eq2.4}
	\text{EDM}(Z_1,Z_2) = \mathop{\lim}\limits_{x \rightarrow \infty } \mathbb{E} \left[ \frac{Z_1}{R} \frac{Z_2}{R} \middle\vert R>x \right]. 
\end{equation}
Based on this relationship, they proposed an estimator for $\text{EDM}(Z_1,Z_2)$, which is defined as
\begin{equation}\label{eq2.5}
	\widehat{\text{EDM}} (Z_1,Z_2) = \frac{1}{N_n} \mathop{\sum}\limits_{i=1}^n \frac{Z_{i1}}{R_i} \frac{Z_{i2}}{R_i} {\textbf{1}_{[R_i \geq x ]}},
\end{equation}
where $\textbf{Z}_i = [Z_{i1},Z_{i2}]^T$ $(1\leq i \leq n)$ is iid random vector, $R_i = \lVert \textbf{Z}_i \lVert $, and $N_n = \mathop{\sum}\limits_{i=1}^n {\textbf{1}_{[R_i \geq x ]}}$. Note that \refeq{eq2.5} suggests the value range of $\text{EDM}$ is $[-0.5, 0.5]$. 
In the next section, we will construct dependence networks among stock returns by using EDM as the main character.

\section{Stock network model based on extremal dependence}\label{sec3}
In this section, we use EDM to construct a network that describes the pairwise extremal dependence structure of stock returns. By specifying such a network structure, we later develop stock selection strategies in Section~\ref{sec4}.
A complex network consists of a set of vertices $V$ and a set of edges $E$, denoted as $G=(V,E)$. An undirected edge connecting vertices $i$ and $j$ is represented as $\{i, j\}$. 
We start by first summarizing important network characteristics, and then discuss how to construct a network using EDMs.

\subsection{The statistical properties of the network}\label{subsec:network}
Complex networks analyze the properties of vertices and edges from a statistical perspective and can describe the characteristics of a network from various aspects. Here we focus on the following six properties and use them in Section~\ref{sec3.2} to compare network characteristics at different thresholds. This analysis will help identify the most suitable threshold for network construction.

\subsubsection{Average degree and degree distribution}\label{sec:degree}
The vertex degree refers to the number of edges connected to a given vertex. The average degree is the mean of the degrees of all vertices in the network and is generally used to measure the overall level of connectivity among vertices. A higher average degree indicates that the edges are more closely connected within the network, suggesting a higher level of interconnection. Degree distribution describes the distribution of degrees among the vertices in the network. If the degree distribution follows a power law, it means that a few vertices have high degrees, while most vertices have lower degrees. Scale-free networks exhibit this property. The degree distribution of a scale-free network is typically represented in a power-law form, as follows (cf. \cite[Chap.1 p.3]{van2024random}) 
\begin{equation}
	\mathbb{P}(n) \propto n^{-\alpha},
\end{equation}
where $\mathbb{P}(n)$ represents the probability density of the \( n \)th vertex, with \( \alpha \) as the estimated parameter.

\subsubsection{Average path length} 
The average path length refers to the mean distance between any two vertices in a network, where distance is typically defined as the minimum number of edges needed to connect the two vertices. A shorter average path length indicates that vertices in the network can influence each other more readily, and information can spread more efficiently across the network. Average path length is a crucial metric for measuring the overall connectivity and efficiency of a network. The calculation formula is as follows (cf. \cite[Chap.1 p.4]{van2024random})
\begin{equation}
	L = \frac{1}{\frac{1}{2}N(N-1)} \sum _{i\geq j}d_{ij},
\end{equation}
where \(N\) denotes the number of vertices, and \(d_{ij}\) represents the number of edges between vertices \(i\) and \(j\).

\subsubsection{Clustering coefficient}
The clustering coefficient measures the degree of clustering or cohesion among vertices in a network. It is defined as the probability that any two neighbors of a given vertex are connected. This is calculated as the ratio of the actual number of connections between neighboring vertices to the maximum possible number of connections between them. The formula to calculate the clustering coefficient is as follows (cf. \cite[Chap.1 p.17]{van2024random})
\begin{equation}
	C_{i} = \frac{2L_i}{k_i (k_i-1)},
\end{equation}
where \(L_i\) represents the actual number of connections between the neighboring vertices of vertex \(i\), and \(k_i\) denotes the number of neighboring vertices for vertex \(i\).

\subsubsection{Network diameter}
The network diameter refers to the maximum distance between any two vertices in a network, where distance is defined as the number of edges that must be traversed to connect the two vertices. Network diameter is defined as (cf. \cite[Chap.1 p.4]{van2024random})
\begin{equation}
	D = \max (d_{ij}).
\end{equation}

\subsubsection{Graph density}
The graph density is the ratio of actual connections in a network to the total possible connections. It reflects the level of connectivity in the network. The calculation formula is as follows (cf. \cite[Chap.1 p.42]{van2024random})
\begin{equation}
	\rho = \frac{M}{\frac{1}{2}N(N-1)},
\end{equation}
where \(M\) represents the actual number of edges in the network.

\subsection{Network construction based on threshold method}\label{sec3.2}
To construct the network of stocks, we denote each stock as a vertex and use a threshold-based approach to construct networks. Here networks constructed under different thresholds have the same number of vertices but differ in the number of edges.  In particular, we define the set of edges $E$ as
\begin{equation}
    E = 
    \left\{ \begin{array}{l} 
    e_{ij}=1,\ \ i\ne j \ \ \text{and} \ \ \text{EDM}(i,j)\ge \theta  \\  
    e_{ij}=0,\ \ \text{otherwise} .
    \end{array}\right.
\end{equation}
In other words, the higher the chosen threshold $\theta$ , the sparser the network.

\subsubsection{Choice of threshold}
When constructing the dependence network between stock returns, one important step is to determine
the $\theta$. Too low a chosen threshold may give a dense network with numerous weak connections, leading to a lack of clear structure. However, a high threshold may overly simplify the network by isolating many vertices such that important connections are missing. Therefore, we need to choose a threshold that preserves important dependence structures, but does not have overwhelmingly high complexity.

Based on the above principles, we set five different threshold values at $\theta = 0.05, 0.1, 0.15, 0.2, 0.25$. For each $\theta$, we construct a corresponding network and examine important characteristics summarized in Section~\ref{subsec:network}. Results are collected in \reftab{tab1}.

\begin{table}[h]
	\centering
	\caption{Comparison of network parameters of different thresholds.}
        \tabcolsep=0.41cm
	\begin{tabular}{ccccccc}
		\toprule
		\multirow{2}{*}{Threshold}& Isolated  & Average  & Network & Graph  & Average clustering  & Average  \\
            & vertex& degree& diameter& density& coefficient&path length\\
        \midrule
		0.05  & 0     & 79.30973 & 0.15605 & 0.70812 & 0.80294 & 0.09323 \\
		0.1   & 2     & 26.63717 & 0.44465 & 0.23783 & 0.53524 & 0.21363 \\
		0.15  & 13    & 5.71681 & 1.61079 & 0.05104 & 0.53920 & 0.61485 \\
		0.2   & 61    & 1.16814 & 1.05244 & 0.01043 & 0.48649 & 0.40297 \\
		0.25  & 98    & 0.19469 & 0.53741 & 0.00174 & 0.75000  & 0.31666 \\
		\bottomrule
	\end{tabular}
	\label{tab1}
\end{table}
We see from Table~\ref{tab1} that the higher the threshold, the more isolated vertices in the network but fewer edges, resulting in lower average degrees. For networks with $\theta = 0.05, 0.01, 0.015$, network diameter, graph density, and average path length all exhibit a positive correlation with the increase in the threshold. The increases in network diameter and average path length reflect reduced connectivity in the network, and the decreasing graph density indicates a sparser network. However, these metrics stop increasing when we further increase $\theta$ to 0.2 and 0.25. The network diameter and path length peak at $\theta=0.15$, but then decline as the number of isolated vertices significantly disrupts the network structure.
Hence, in the sequel, we do not further consider networks with thresholds ranging from 0.2 to 0.25.
\begin{figure}[h]
	\centering
	\subcaptionbox{Threshold=0.05}{
		\begin{minipage}[h]{.45\linewidth}
			\centering
			\includegraphics[scale=0.4]{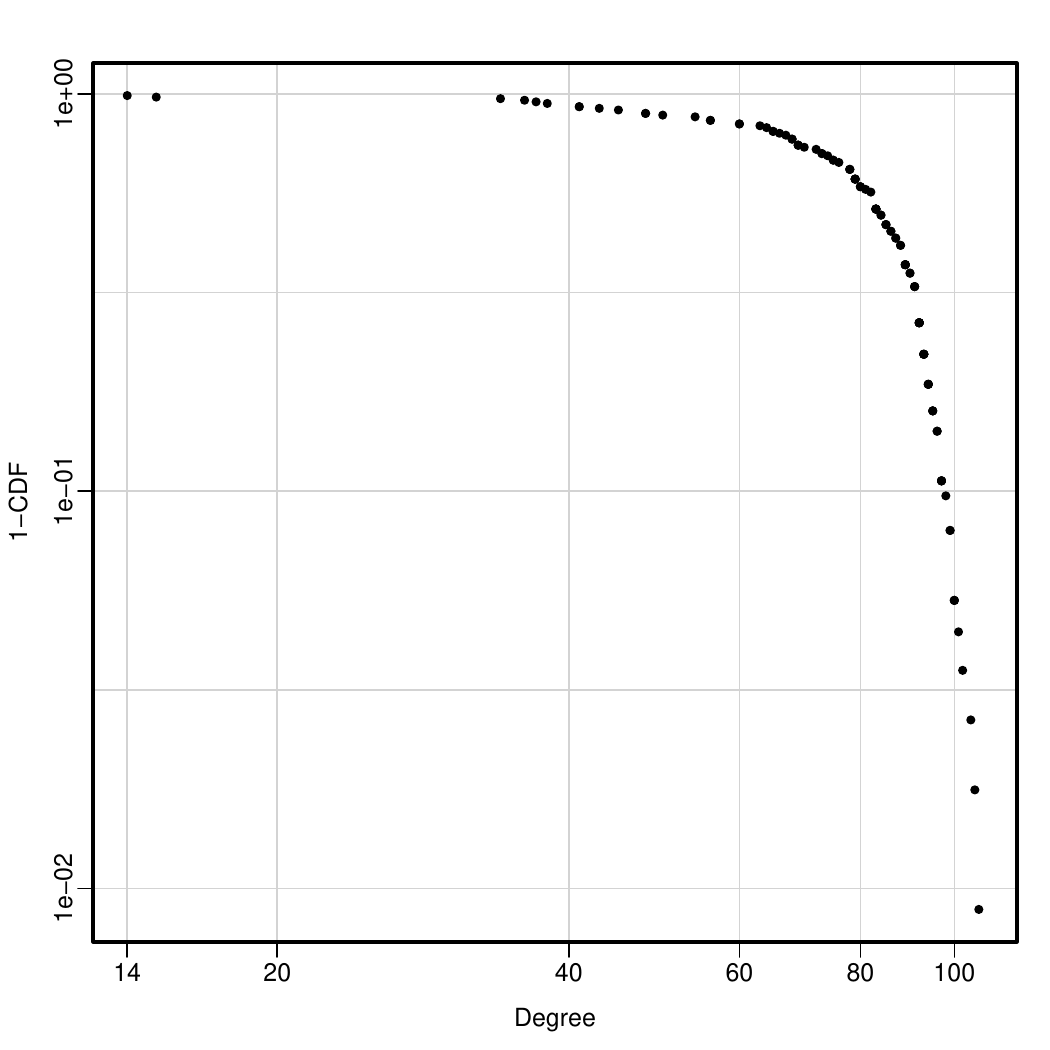}
		\end{minipage}
	}
	\subcaptionbox{Threshold=0.1}{
		\begin{minipage}[h]{.45\linewidth}
			\centering
			\includegraphics[scale=0.4]{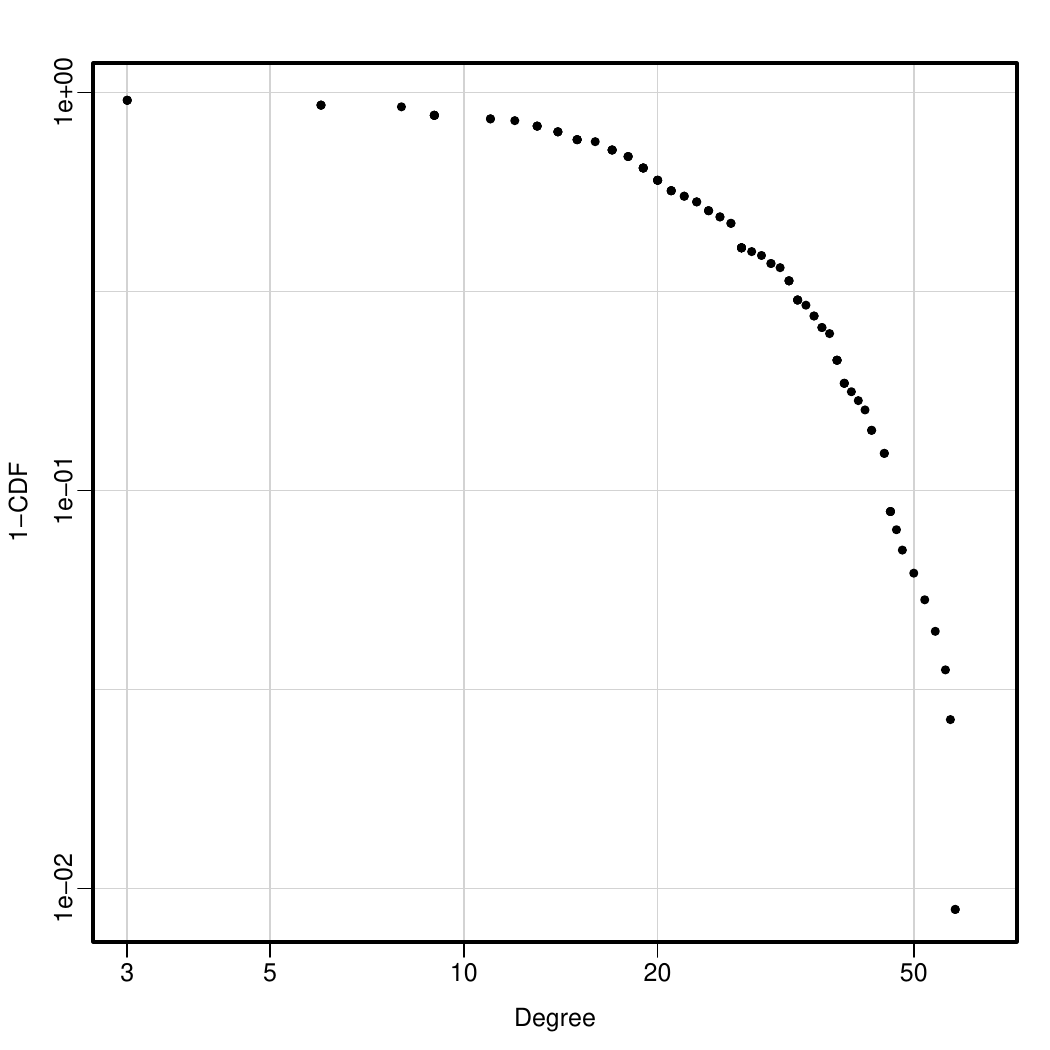}
		\end{minipage}
	}\\
	\subcaptionbox{Threshold=0.15}{
		\begin{minipage}[h]{.45\linewidth}
			\centering
			\includegraphics[scale=0.4]{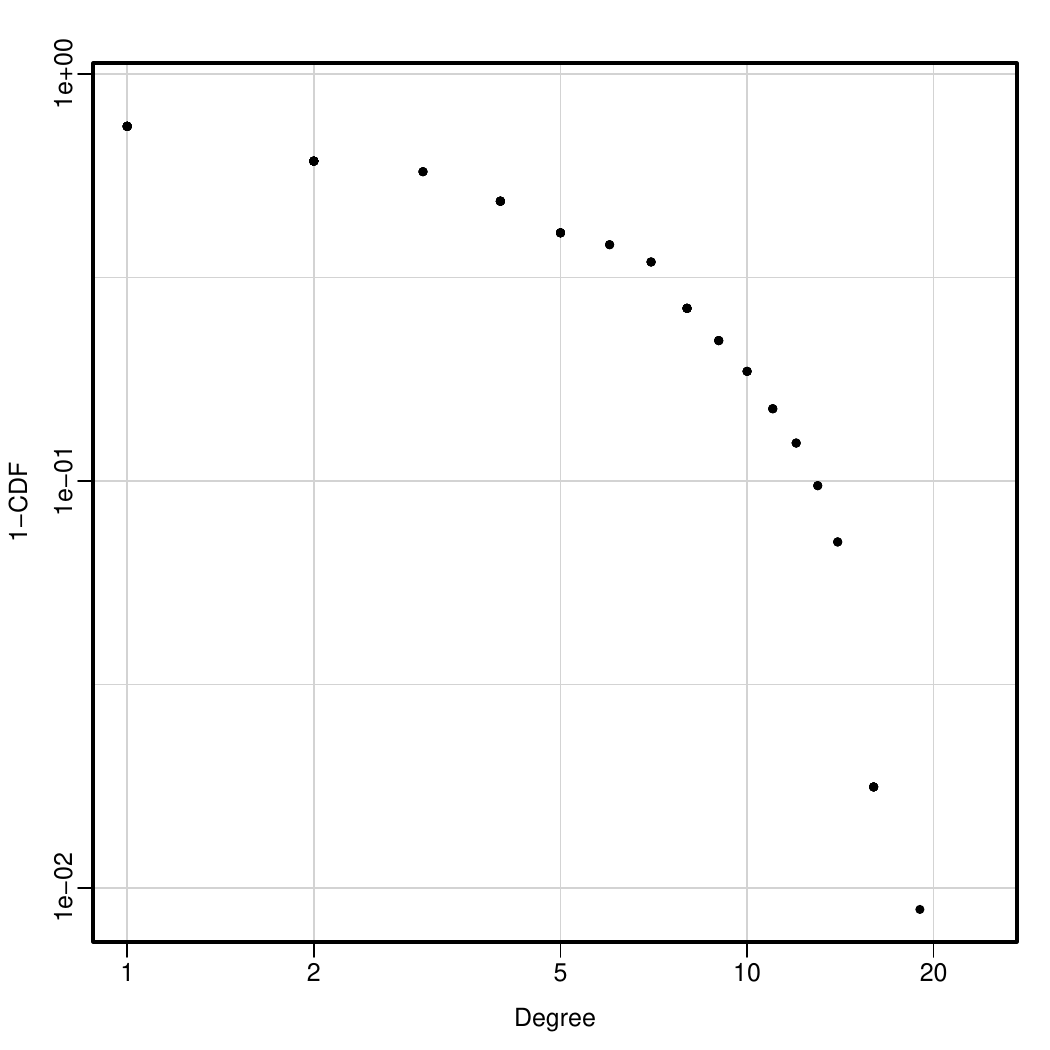}
		\end{minipage}
	}
	\subcaptionbox{Threshold=0.2}{
		\begin{minipage}[h]{.45\linewidth}
			\centering
			\includegraphics[scale=0.4]{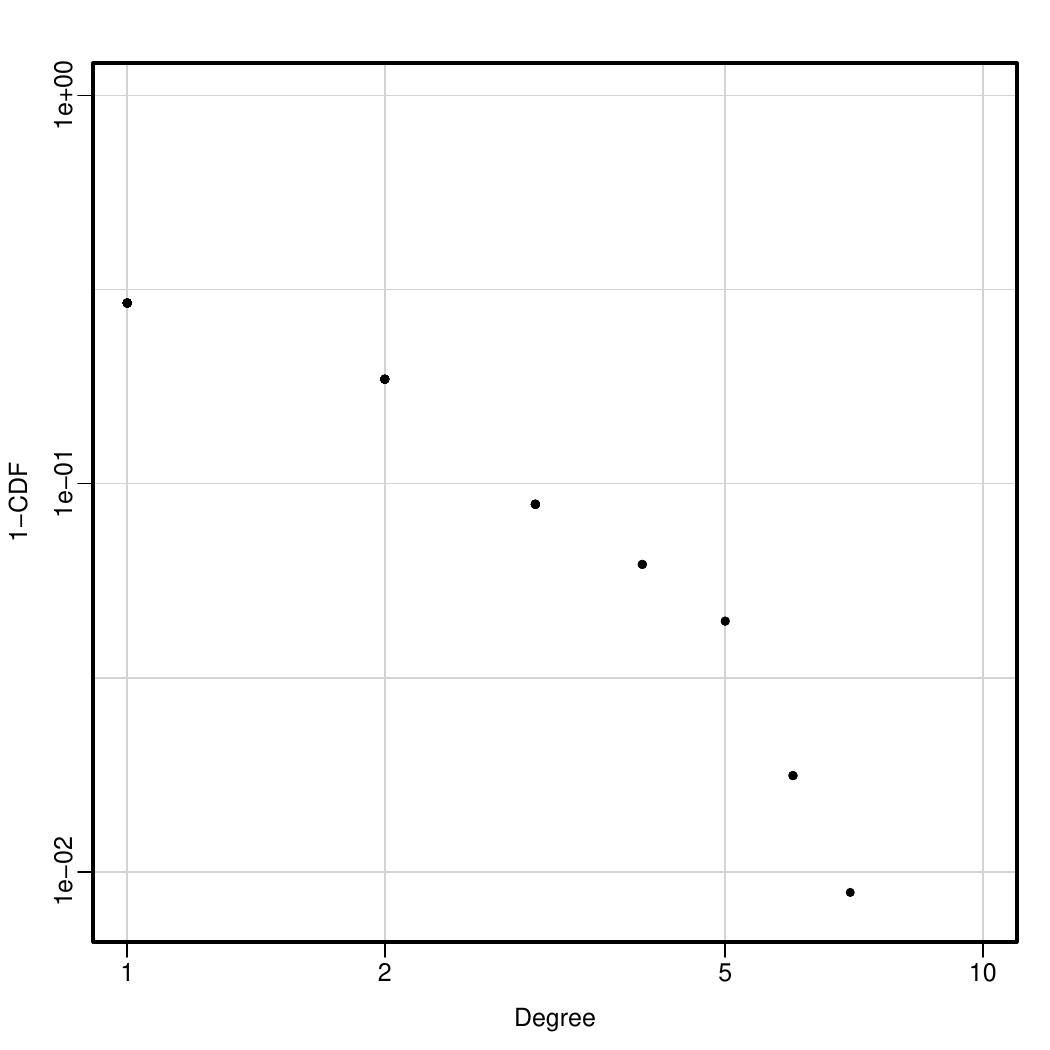}
		\end{minipage}
	}
	\caption{Plots of the complementary cumulative distribution function (1-CDF) for the degrees corresponding to different thresholds on a log-log scale.}
	\label{fig:threshold}
\end{figure}

\begin{figure}[h]
	\centering
	\includegraphics[scale=0.8]{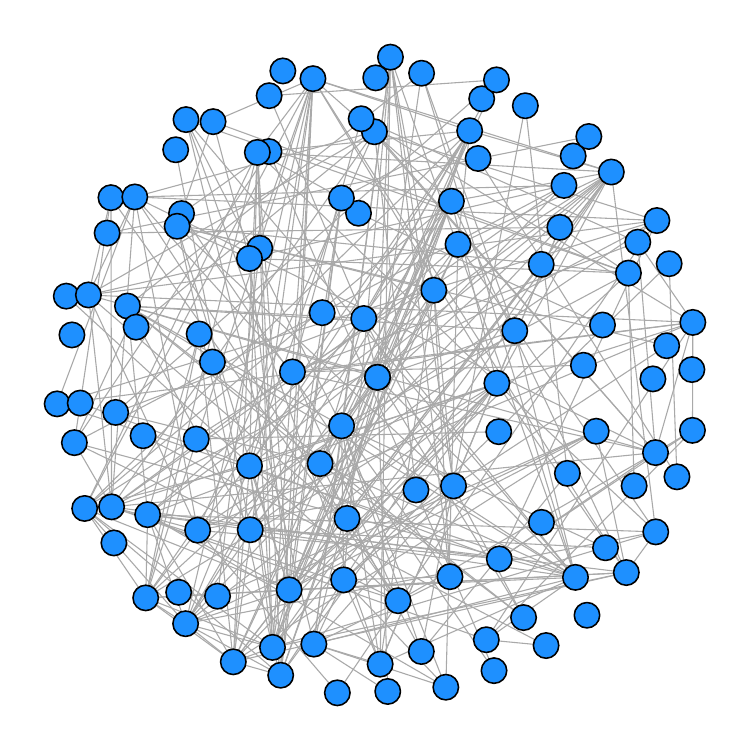}
	\caption{A dependence network whose threshold is 0.15.}
	\label{fig:network}
\end{figure}

For thresholds ranging from 0.05 to 0.2, we also analyze their empirical degree distributions. The complementary cumulative distribution function (1-CDF) plots of degree values are given in \reffig{fig:threshold}.
\reffig{fig:threshold}(a) and \reffig{fig:threshold}(b) correspond to the empirical degree distribution of two networks with $\theta = 0.05, 0.1$, respectively. These two plots exhibit rapid decay, with only a few vertices retaining high degrees and most vertices keeping a much lower degree, showing little evidence of the scale-free property (cf. Section~\ref{sec:degree}). 
However, for $\theta = 0.15$, \reffig{fig:threshold}(c) reflects a power-law decay pattern for the degree tail distribution. Therefore, we proceed by setting $\theta = 0.15$ and the network is visualized in \reffig{fig:network}, which contains 113 vertices. When choosing $\theta = 0.15$, the value of EDM is ranging from $-0.041$ to $0.5$.

In what follows, we refer to such a graph as the \emph{dependence network} for stocks, and wherever no edge has been observed between nodes $i$ and $j$, it means the corresponding two stock returns show asymptotic independence. We will further analyze the properties of this network in Section~\ref{subsec:portfolio}.

\subsubsection{Maximum independent set}
After constructing the dependence network as in \reffig{fig:network}, we now aim to propose our portfolio strategy by identifying a collection of stocks whose returns are exhibiting low extremal dependence. Then the corresponding portfolio should perform well against extremal risk in the market. We achieve this goal by finding the \emph{maximum independent set} of the network.

For an undirected graph $G=(V,E)$, if $V^*\subseteq V$ and any two vertices in $V^*$ are not connected, then $V^*$ forms an independent set in graph $G$. If $V^*$ is not contained in any other independent set, it is called a maximal independent set. If the size of $V^*$ is the largest among all maximal independent sets, it is referred to as the maximum independent set. The maximum independent set problem (MISP) is a classic combinatorial optimization problem in graph theory. 

Finding an exact solution to MISP for a given graph has been shown to be NP-hard (cf. \cite{karp2010reducibility}). Therefore, as the size of the graph increases, the time complexity of solving MISP also increases, rendering exact solutions impractical. As a result, many researchers have developed heuristic-based approximate algorithms to solve the MISP (cf. \cite{palubeckis2008recursive, peng2015performance, wu2012multi}). Although these algorithms cannot guarantee optimal solutions, they can search much faster and guarantee cost-effectiveness when solving large-scale MISP problems. Currently, some of the most widely used heuristic algorithms include the greedy algorithm \cite{palubeckis2008recursive}, local search \cite{peng2015performance}, and Tabu search \cite{wu2012multi}. 
In this paper, we use the greedy algorithm in \cite{palubeckis2008recursive} to find solutions to the MISP. This algorithm ensures that the solution set we find is an independent one by gradually expanding the vertex set until all possibilities are exhausted to obtain a feasible solution. 

\subsection{Stock portfolio strategy based on complex networks}\label{subsec:portfolio}
We now propose our graph-based portfolio strategy by first partitioning the entire network with both sector-based and community-based classification methods. Then for each sub-network, we apply graph-theoretical algorithms to find the corresponding MISP, which gives the selected portfolio. To evaluate the effectiveness of this strategy, we use both value at risk (VaR) and expected shortfall (ES) as risk measurements; they provide valuable insights for investors to avoid extremal risk.

\subsubsection{Sector-based classification}
Based on the ``CSI Industry Classification Standard Description" released by China Securities Index Co., Ltd. in December 2021, the 113 Shenzhen component stocks are classified into 11 primary industry categories: information technology, industrials, healthcare, consumer staples, materials, communication services, financials, consumer discretionary, real estate, utilities, and energy. The number of stocks in each sector is summarized in \reftab{tab4}.
\begin{table}[H]
	\centering
	\caption{The number of stocks corresponding to each industry sector.}
	\tabcolsep=0.85cm
	\begin{tabular}{cccc}
		\toprule
		Sector  & Number   & Sector  & Number \\
		\midrule
		Information technology  & 25    & Financials & 7 \\
		Industrials    & 24    & Consumer discretionary & 7 \\
		Healthcare  & 16    & Real estate & 3 \\
		Consumer staples  & 11    & Utilities & 2 \\
		Materials   & 10    & Energy & 1 \\
		Communication services  & 7     &       &  \\
		\bottomrule
	\end{tabular}
	\label{tab4}
\end{table}

\begin{figure}[h]
	\centering
	\includegraphics[scale=0.7]{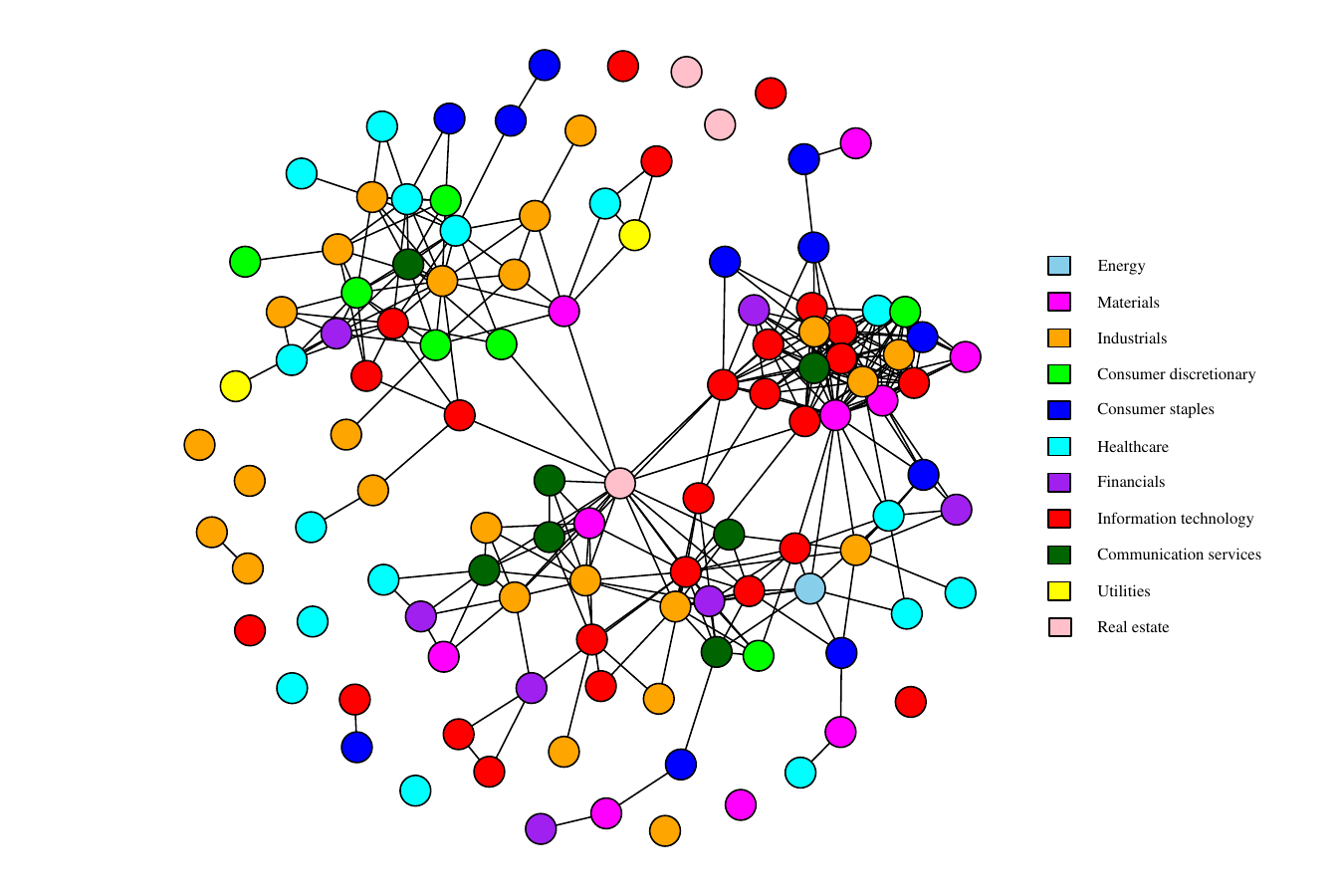}
	\caption{A stock network is classified into 11 industry sectors, with each sector represented by a different color.}
	\label{fig:graphindustry}
\end{figure}

In \reftab{tab4}, information technology, industrials, and healthcare sectors rank as the top three in terms of the number of stocks they contain, indicating that these three sectors play a significant role in the Chinese financial market. Given that companies in the same sector partake in similar business activities and maintain comparable relationships with companies from other sectors, it is reasonable to anticipate that these stocks may demonstrate similar dependencies and be grouped into the same cluster. In Figure~\ref{fig:graphindustry}, we visualize the network, where vertices are colored based on the sectors they belong to.

\begin{figure}[h]
	\centering
	\subcaptionbox{Information technology}{
		\begin{minipage}[h]{.3\linewidth}
			\centering
			\includegraphics[scale=0.3]{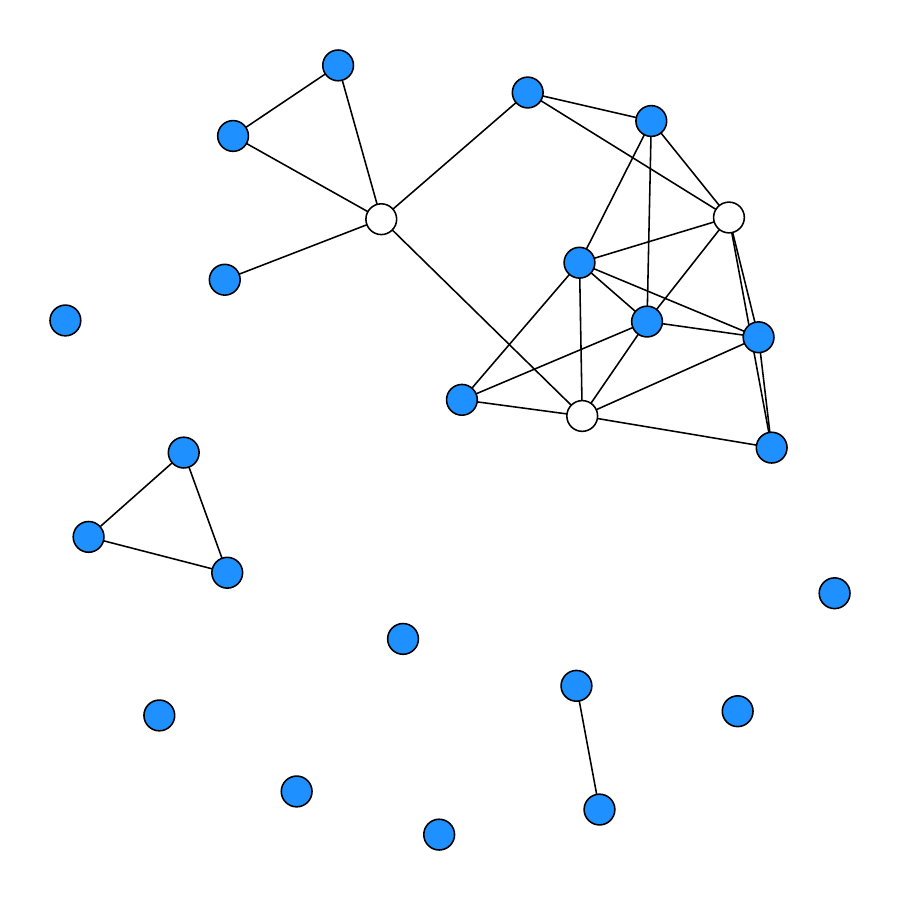}
		\end{minipage}
	}
	\subcaptionbox{Industrials}{
		\begin{minipage}[h]{.3\linewidth}
			\centering
			\includegraphics[scale=0.3]{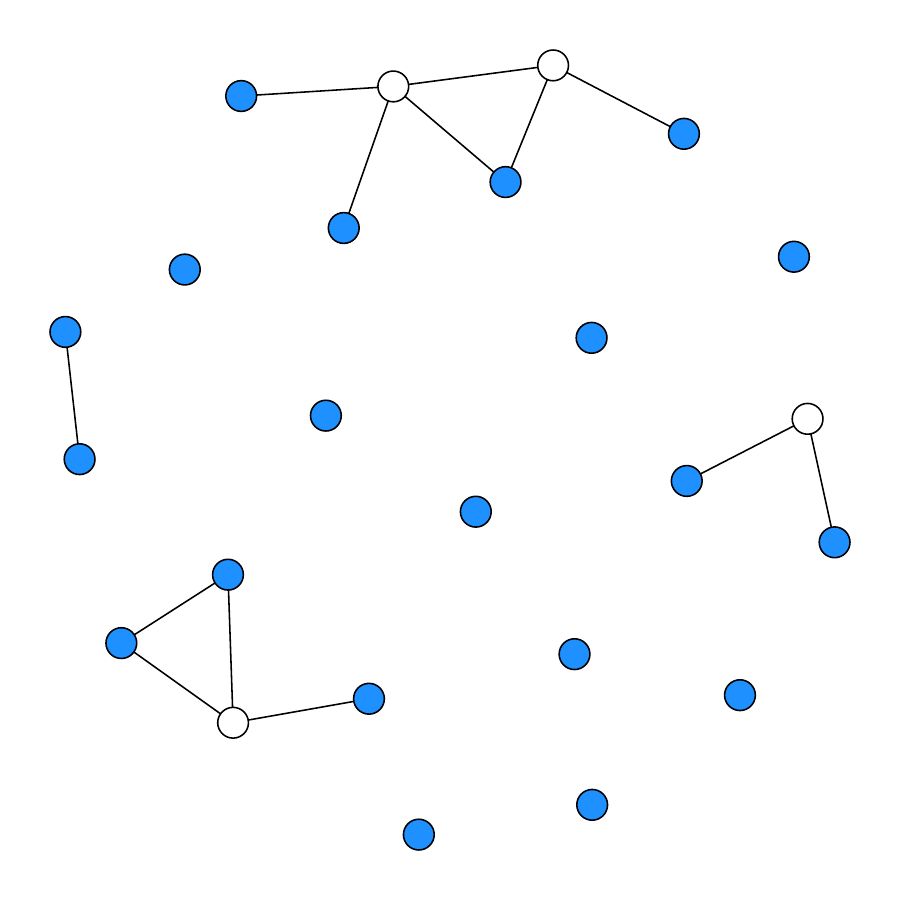}
		\end{minipage}
	}
	\subcaptionbox{Healthcare}{
		\begin{minipage}[h]{.3\linewidth}
			\centering
			\includegraphics[scale=0.3]{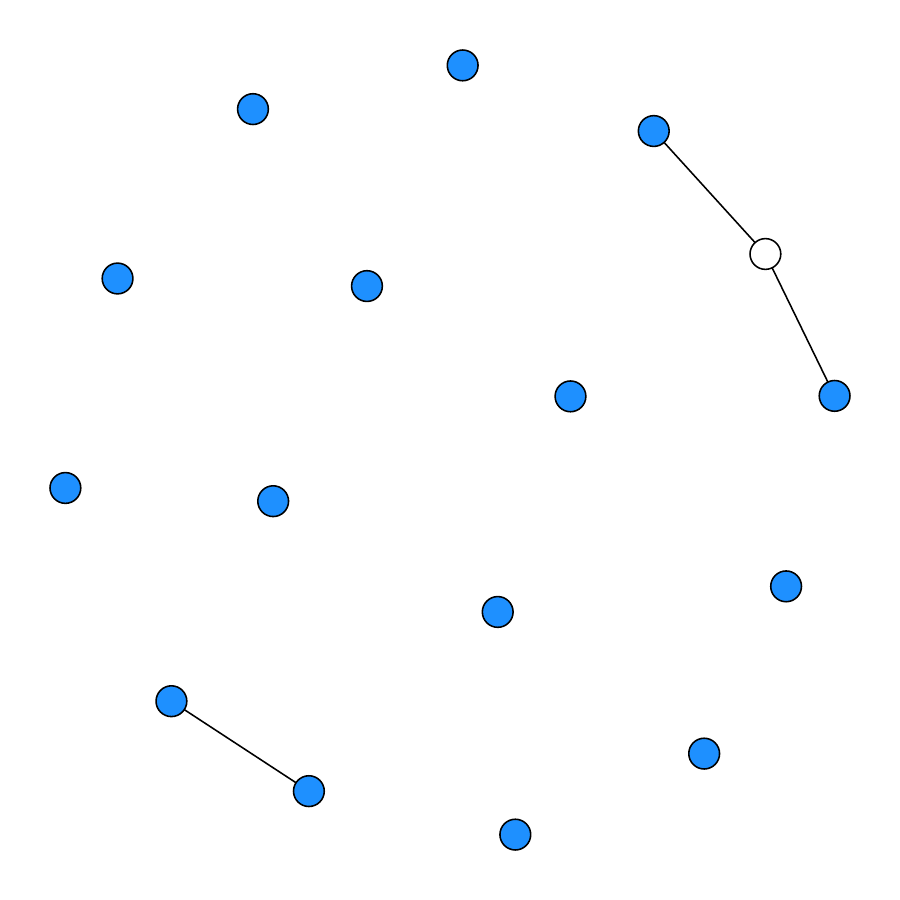}
		\end{minipage}
	}
	\caption{Sub-networks extracted from the original network; the blue nodes represent the maximum independent set of results of the sub-network by sector.}
	\label{fig:dep}
\end{figure}

In \reffig{fig:graphindustry}, red, orange, and cyan represent the stocks in the Information Technology, Industrials, and Healthcare sectors, respectively. Contrary to the assumption that stocks within the same sector tend to cluster, in most cases, stocks from the same sector do not belong to one single cluster. Following such sector-based classification of the dependence network, the maximum independent set for each sector is obtained and summarized in \reffig{fig:dep}.

\reffig{fig:dep} reveals that after extracting sub-networks, a large number of isolated vertices appear, whereas the original network only contains 13 isolated vertices. This phenomenon suggests strong extremal dependence among different sectors. Consequently, when one sector is forcibly removed from the whole, the connections between the different sectors are ignored.
Later in Section~\ref{subsec:overall}, we find that when considering the overall market, the sector-based portfolio has more volatile returns and risks than the community-based one, so we conclude that partitioning the dependence network by sectors is suboptimal.


\subsubsection{Community detection of the dependence network based on GN algorithm}\label{subsec:community}
\begin{figure}[h]
	\centering
	\includegraphics[scale=0.4]{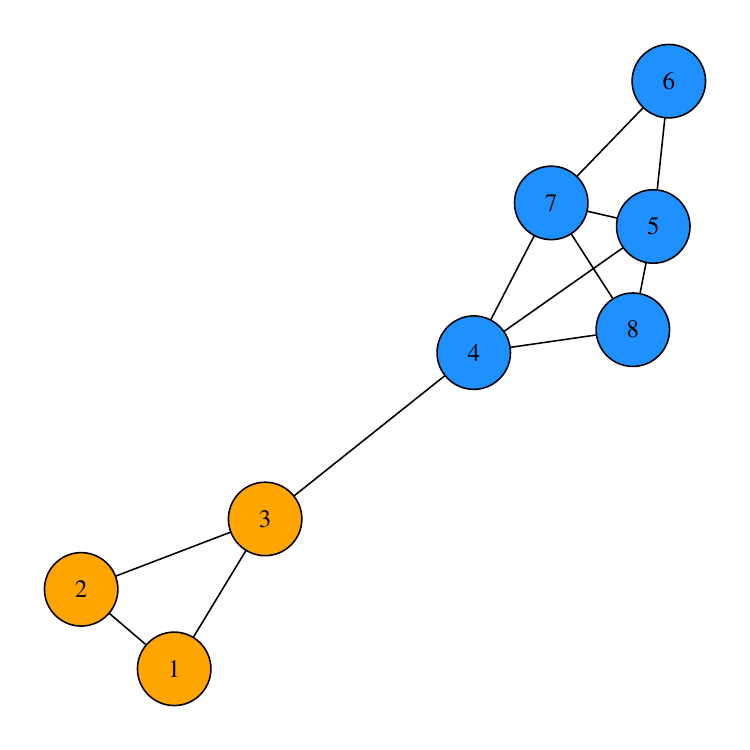}
	\caption{A network diagram with 2 communities in different colours.}
	\label{fig:example}
\end{figure}

To compare with the sector-based portfolio construction, we now use a community-based strategy to build the portfolio. Community structure is a key feature of complex networks, where certain vertices naturally group together, forming smaller, densely connected communities, as shown in \reffig{fig:example}. Within these communities, internal connections are dense, while connections between communities are sparse, revealing a modular organization. This suggests that vertices within the same community have close relationships, likely due to similar characteristics or roles. The concept of community was first introduced in sociology, and it has since found extensive applications across various disciplines, including physics, biology, electronics, and computer science (cf. \cite{bazzi2016community, fortunato2016community, isogai2014clustering, isogai2017dynamic}).

\begin{figure}[h]
	\centering
	\includegraphics[scale=0.6]{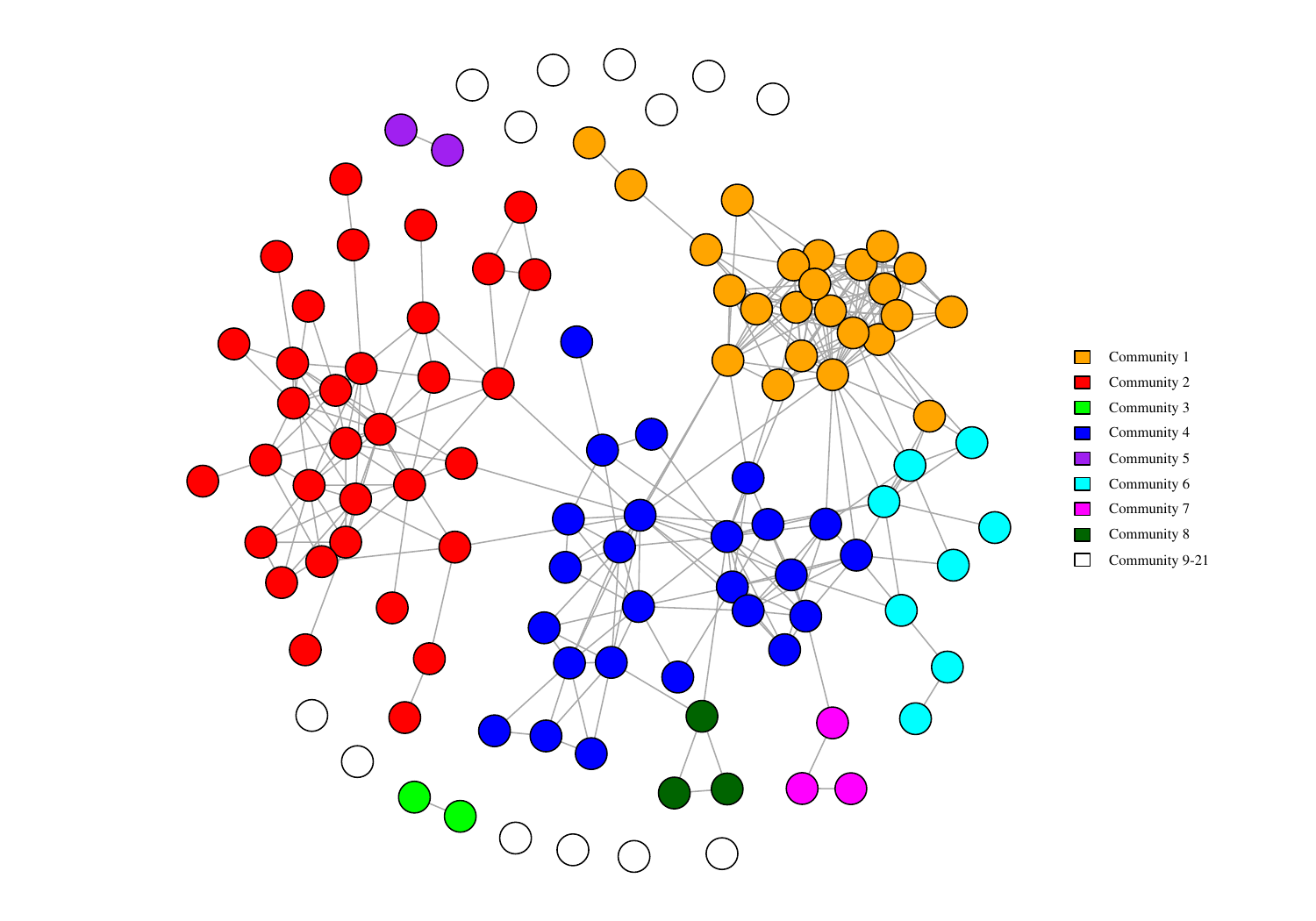}
	\caption{The stock network consists of 13 communities, with each community represented by a different color.}
	\label{fig:gcomm}
\end{figure}

Currently, there are numerous algorithms available for identifying community structures in networks (cf. \cite{kernighan1970efficient, capocci2005detecting, pothen1990partitioning, clauset2004finding, newman2004fast}). In what follows, we use the Girvan-Newman algorithm to partition the network into 21 communities, 13 of which consist of only a single vertex (i.e. isolated vertices). 
The graph with communities distinguished by different colors is shown in \reffig{fig:gcomm}.

\begin{figure}[h]
	\centering
	\subcaptionbox{Community 1}{
		\begin{minipage}[h]{.3\linewidth}
			\centering
			\includegraphics[scale=0.23]{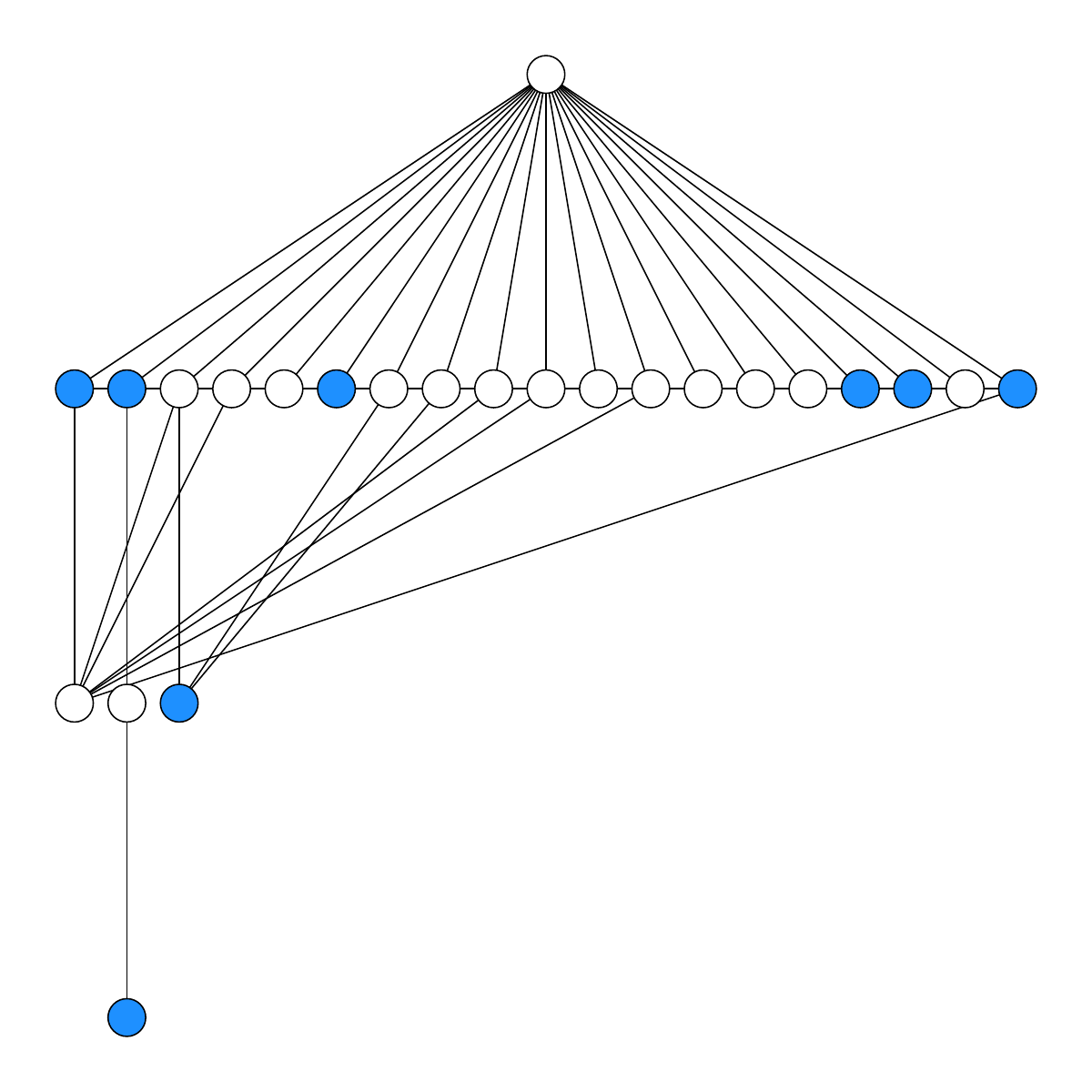}
		\end{minipage}
	}
	\subcaptionbox{Community 2}{
		\begin{minipage}[h]{.3\linewidth}
			\centering
			\includegraphics[scale=0.26]{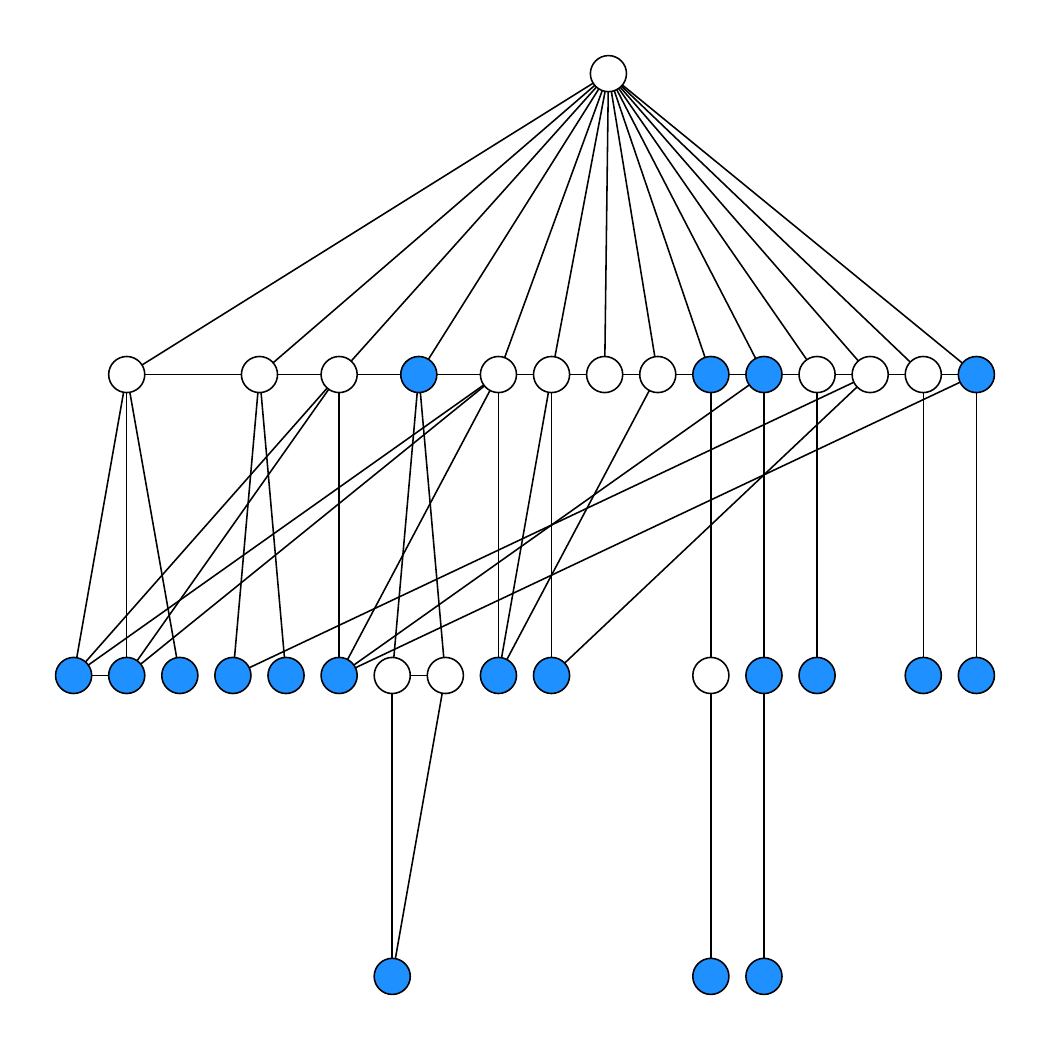}
		\end{minipage}
	}
	\subcaptionbox{Community 4}{
		\begin{minipage}[h]{.3\linewidth}
			\centering
			\includegraphics[scale=0.26]{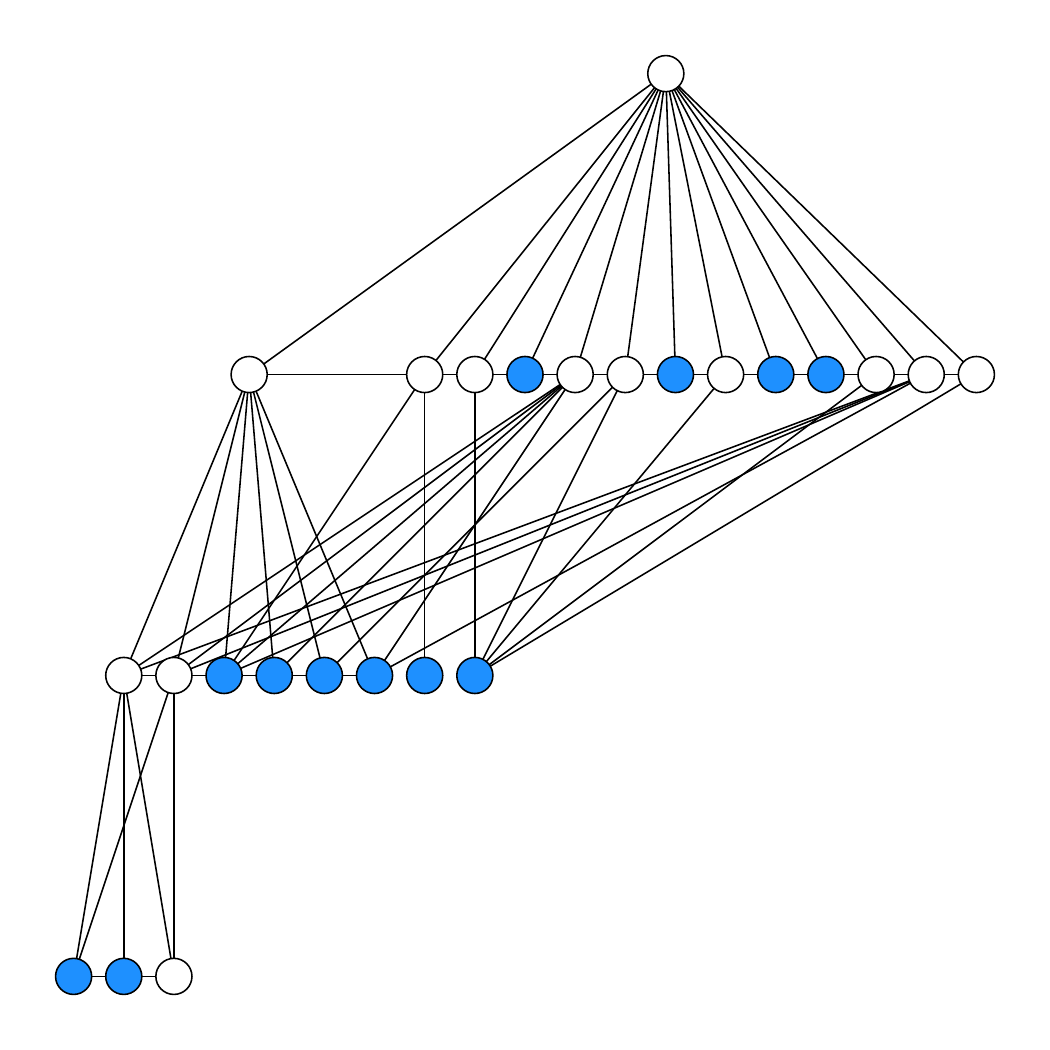}
		\end{minipage}
	}
	\caption{Sub-networks extracted from the original network; the blue nodes represent the maximum independent set of results of the sub-network by community.}
	\label{fig:commu}
\end{figure}

Comparing to \reffig{fig:graphindustry}, where the vertices exhibit a disordered color distribution, \reffig{fig:gcomm} with vertices colored based on different communities, displays a well-organized pattern, with vertices of the same color predominantly clustered within the same community. This graph demonstrates a stronger clustering structure. We also see that communities 1, 2, and 4 account for the largest proportion in the entire network, with a total share of approximately $80\%$. Moreover, all influential industries have stocks within these three key communities. Therefore, we focus on these three key communities, and find the maximum independent set of each community. Results for communities 1, 2, and 4 are summarized in \reffig{fig:commu}, where nodes in blue represent the maximum independent set.

\section{Empirical study and results}\label{sec4}
In this section, we propose a portfolio strategy to minimize the risk of extremal loss, where two common risk measures, value-at-risk (VaR) and expected shortfall (ES) are used. Additionally, we compare local and overall portfolios, offering investment advice tailored to different levels of risk tolerance.

\subsection{Optimal portfolio with minimum risk}
VaR \cite{sunaryo2007manajemen} quantifies the market risk of a stock investment portfolio, and the VaR at the confidence level of $1 - \alpha$ is computed by
\begin{equation}
    \mathbb{P} (\Delta P < -{\text{VaR}}) = \alpha ,
\end{equation}
where $\Delta P = P(t+\Delta t) - P(t)$ represents the loss of the investment portfolio over the holding period of length $\Delta t$.

In fact, a \emph{coherent} risk measure satisfies the following four axioms: translation invariance, subadditivity, positive homogeneity, and monotonicity, and VaR is not coherent since it fails to satisfy the subadditivity property (cf. \cite{ES}).
Instead, the expected shortfall (ES) \cite{ES} defined as
\begin{equation}
    \label{eq:def_ES}
	{\text{ES}} = \mathbb{E}[L|L>{\text{VaR}}]
\end{equation}
is proven to be coherent. VaR specifies the maximum expected loss that will not be exceeded in the absence of adverse events, while ES quantifies the expected loss one may face in the event of actual adverse scenarios. In what follows, we analyze and select investment portfolios with different weights, using VaR and ES as risk measures.

We assume the holding period $\Delta t$ is 1 day and
calculate the VaR and ES of each stock at a $95\%$ confidence level. The objective function is to minimize the overall risk of the portfolio. Constraints are imposed to ensure that the sum of weights equals 1, with each weight coefficient ranging from 0 to 0.3. Since the current three-month deposit rate set by the Chinese Central Bank is $1.15\%$, we restrict the desired overall return rate to be at least $1.15\%$. This problem is formulated as a linear programming problem:
\begin{equation}\label{eq:portfolio}
    \begin{split}
        &\min \ \ \sum\limits_{i=1}^n c_i Risk_i \\
        &s.t. \left\{\begin{array}{lc}
            & \sum\limits_{i=1}^n c_i=1 \\
            & 0\leq c_i \leq 0.3 \\
            & \sum\limits_{i=1}^n c_iR_i \geq 1.15\%,
        \end{array}\right.
    \end{split}
\end{equation}
where $Risk_i$ refers to the associated VaR or ES of stock $i$, $c_i$  denotes the weight of each stock, and $R_i$ denotes the return of stock $i$. 

\subsection{Local portfolio analysis}
For illustration purposes, we take community 1 and the sector healthcare as representative examples to analyze the performance of the local portfolio strategy. We solve the minimization problem in \eqref{eq:portfolio} by using the \verb6linprog6 function in MATLAB and results are collected in Tables~\ref{tab:var4} -- \ref{tab:weightindep}.

\begin{table}[h]
	\centering
	\caption{The VaR and ES of the maximum independent set in community 1 (at a confidence level of 95\% ).}
	\tabcolsep=0.75cm
	\begin{tabular}{ccccc}
		\toprule
		& Jinlang & Leading & \multirow{2}{*}{Zoomlion}& \multirow{2}{*}{Yanghe Stock}\\
            & Technology& Intelligence& &\\
		\midrule
		VaR (\%)   & 4.3969& 3.1210& 1.7268& 2.3910\\
		ES (\%)    & 5.3046& 3.5588& 2.4963& 2.9918\\
  	\bottomrule
		\toprule
		& \multirow{2}{*}{Bettenie}& Zhifei & Yiling & 37 Mutual\\
            & & Biology& Pharmaceutical&Entertainment\\
		\midrule
		VaR (\%)   & 3.5109& 2.9358& 3.0133& 5.3156\\
		ES (\%)    & 5.2481& 4.3478& 3.9266& 7.6437\\
		\bottomrule
	\end{tabular}
	\label{tab:var4}
\end{table}

\begin{table}[h]
	\centering
	\caption{The VaR and ES of the maximum independent set in the healthcare sector (at a confidence level of 95\% ).}
	\tabcolsep=0.2cm
	\begin{tabular}{cccccc}
		\toprule
		& Yunnan & Huadong& Sanjiu Medical& \multirow{2}{*}{NHU}& Hualan Biological\\
            & Baiyao& Medicine& \&Pharmaceutical& & Engineering\\
		\midrule
		VaR (\%)& 1.7355& 2.4906& 2.9746& 1.8171& 2.1842\\
		ES (\%)& 2.6704& 3.6875& 4.4252& 2.3141& 2.8016\\
  	\bottomrule
        \toprule
         & RAAS& Yiling & Aier Eye& Zhifei& \\
         & Blood& Pharmaceutical& Hospital& Biological& \\
        \midrule
        VaR (\%)& 2.0311& 3.0085& 3.0468& 2.9358& \\
        ES (\%)& 2.5940& 3.9266& 3.6342& 4.3478& \\
        \bottomrule
		\toprule
		 & Walvax& Tigermed& Pharmaron& Mindray Bio-Medical& \\
         & Biotechnology& Consulting& Beijing& Electronics& \\
		\midrule
		VaR (\%)   & 2.8846& 4.1350& 4.0764& 2.4367& \\
		ES (\%)    & 3.7490& 5.5424& 6.8838& 3.4101& \\
		\bottomrule
	\end{tabular}
	\label{tab:varindep}
\end{table}

For both community 1 and the healthcare sector, we first compute their maximum independence sets and report the corresponding values (with a $95\%$ confidence level) in \reftab{tab:var4} and \reftab{tab:varindep}, respectively. The ES values for each portfolio are consistently larger than those of VaR since ES refers to the loss expectation under the condition that the loss exceeds VaR (cf. \eqref{eq:def_ES}). 
Next, we solve the optimization problem in \eqref{eq:portfolio} to obtain the optimal portfolio weights for each maximum independence set, and results are summarized in \reftab{tab:weight4} and \reftab{tab:weightindep}. 

\begin{table}[h]
	\centering
	\caption{Optimal portfolio with the minimum risk for the maximum independent set in community 1; the obtained objective function values are 2.43\% (VaR) and 3.24\% (ES), and the total return rate is 1.15\% for both cases.}
	\tabcolsep=0.67cm
	\begin{tabular}{ccccc}
		\toprule
		& Jinlang & Leading & \multirow{2}{*}{Zoomlion}& \multirow{2}{*}{Yanghe Stock}\\
            & Technology& Intelligence& &\\
		\midrule
		Weight (VaR)& 0& 0& 0.3000    & 0.3000    \\
		Weight (ES)& 0& 0.0261& 0.3000    & 0.3000    \\
  	\bottomrule
		\toprule
		& \multirow{2}{*}{Bettenie}& Zhifei & Yiling & 37 Mutual\\
            & & Biology& Pharmaceutical&Entertainment\\
		\midrule
		Weight (VaR)& 0& 0.1779& 0.2221& 0\\
		Weight (ES)& 0& 0.0739& 0.3000& 0\\
		\bottomrule
	\end{tabular}
	\label{tab:weight4}
\end{table}

\begin{table}[h]
	\centering
	\caption{Optimal portfolio with the minimum risk for the maximum independent set in the healthcare sector; the obtained objective function values are 2.09\% (VaR) and 2.96\% (ES), and the total return rate is 1.15\% for both cases.
}
	\tabcolsep=0.2cm
	\begin{tabular}{cccccc}
		\toprule
		& Yunnan & Huadong& Sanjiu Medical& \multirow{2}{*}{NHU}& Hualan Biological\\
        & Baiyao& Medicine& \&Pharmaceutical& & Engineering\\
		\midrule
		Weight (VaR)& 0.3000& 0.3000& 0& 0.3000& 0.0196\\
		Weight (ES)& 0.3000& 0.3000& 0& 0.3000& 0.0196\\
  	\bottomrule
        \toprule
         & RAAS& Yiling & Aier Eye& Zhifei& \\
         & Blood& Pharmaceutical& Hospital&Biological& \\
        \midrule
        Weight (VaR)& 0& 0& 0& 0& \\
        Weight (ES)& 0& 0& 0& 0& \\
        \bottomrule
		\toprule
		 & \multirow{2}{*}{Walvax} & Huichuan & Mindray & Huali & \\
         & & Tigermed& Pharmaron& Mindray Bio-Medical& \\
		\midrule
		Weight (VaR)& 0.0804& 0& 0& 0& \\
		Weight (ES)& 0.0804& 0& 0& 0& \\
		\bottomrule
	\end{tabular}
	\label{tab:weightindep}
\end{table}

From Tables~\ref{tab:weight4} and \ref{tab:weightindep}, we see that for both community 1 and the healthcare sector, either using VaR or ES as the risk measure will give similar decisions on which stocks should not be invested in the portfolio. 
Furthermore, in terms of the optimal portfolio weight, no matter which risk measure is used, \reftab{tab:weightindep} reports the same weights for each selected stock. 
However, as shown in \reftab{tab:weight4}, when VaR is used as the risk measure, our proposed strategy suggests to invest in Zhifei Biology more. On the other hand, if ES is employed in the objective function of \eqref{eq:portfolio}, the weight in Zhifei Biology is significantly reduced, while stocks like Leading Intelligence and Yiling Pharmaceutical receive a larger allocation. Such discrepancy arises because the ES for Zhifei Biology is much larger than its VaR, even exceeding those of Leading Intelligence and Yiling Pharmaceutical.

\begin{figure}[h]
	\centering
	\begin{subfigure}[t]{1\linewidth}
		\centering
		\includegraphics[scale=0.48]{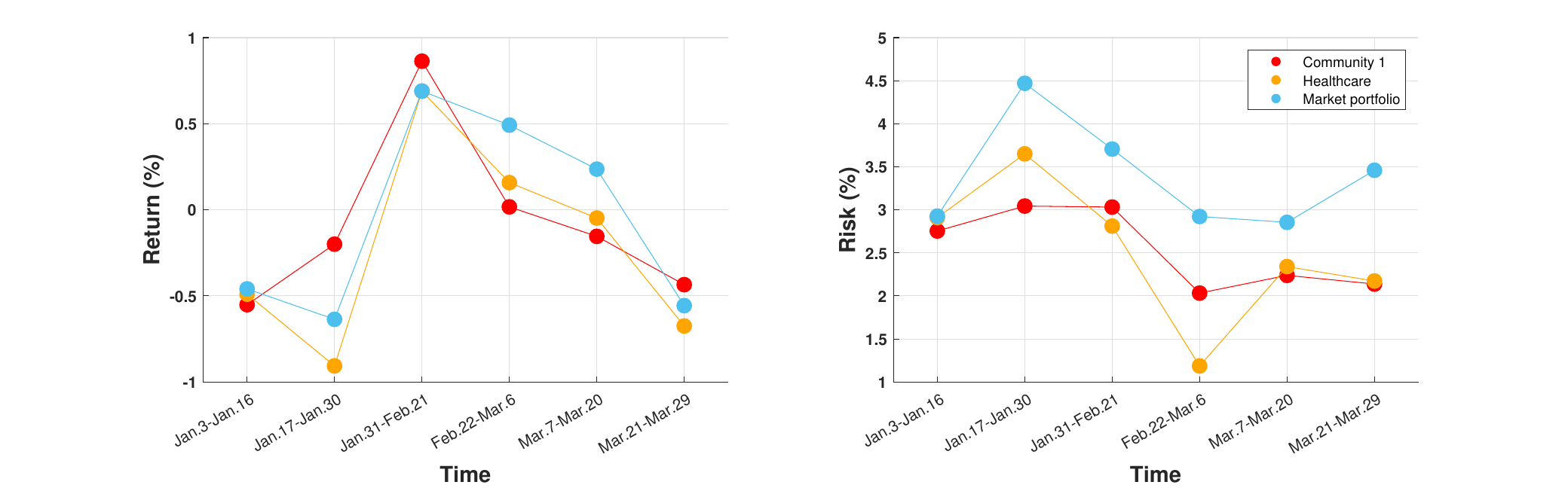}
		\caption{VaR as the risk measure}
	\end{subfigure}
	\begin{subfigure}[b]{1\linewidth}
		\centering
		\includegraphics[scale=0.48]{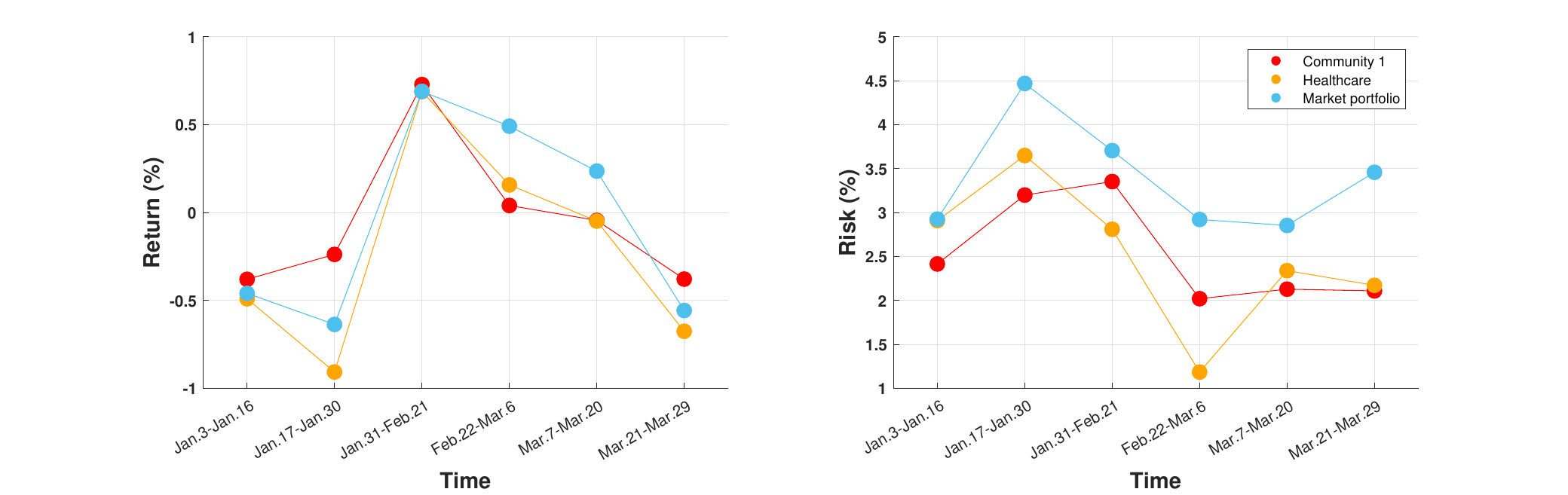}
		\caption{ES as the risk measure}
	\end{subfigure}
	\caption{Line plots of the return-time (left) and risk-time (right) with different risk measure by VaR (a) and ES (b); the red, yellow, and blue nodes respectively represent the maximum independent set in community 1, healthcare sector and the market portfolio.}
	\label{fig:pp}
\end{figure}

To further assess the performance of our stock portfolios in 2024, we obtain stock prices from January 1, 2024, to March 31, 2024, and segment it into six intervals of 10 trading days each. We compare the actual portfolio returns with the market portfolio (Shenzhen component of the CSI 300), 
and results are presented in \reffig{fig:pp}.

In the above panels of \reffig{fig:pp}, we plot the returns and risks of the three different portfolios suggested by the maximum independence set of community 1 (red) and the healthcare sector (yellow) as well as the overall market index (blue). To make detailed comparisons, we further give the line plots of ES as a risk measure in \reffig{fig:pp}(b). 

First, no matter which risk measure is used, we see that the proposed strategy based on the maximum independence set can reduce the portfolio risk for both community 1 and the healthcare sector over all time windows, compared to that of the chosen market portfolio. This confirms the effectiveness of our strategy against the extremal risk. In particular, from the return plots, we observe that especially when the market portfolio experiences a huge loss (e.g. Jan 17-30, 2024), the portfolio constructed from community 1 has a much better performance (higher return and lower risk) than the market portfolio. 

\subsection{Overall portfolio analysis}\label{subsec:overall}
Since the maximum independent set demonstrates superior return-risk performance compared to the market portfolio, we further extend such a local portfolio strategy to the global scale. However, given that the maximum independent set problem is NP-hard, solving it for the entire network is computationally infeasible. To address this challenge, we first use sector-based and community-based classifications to identify the maximum independent sets within the network and then derive an overall optimal portfolio. 

Starting with the sector-based classification, we solve the maximum independent set for each of the 11 sectors individually and then aggregate the obtained sets to address the optimization problem in \eqref{eq:portfolio}. 
\begin{table}[h]
	\centering
	\caption{Optimal portfolio with the minimum risk for the sector-based overall stocks; the obtained objective function values are 1.75\% (VaR) and 2.43\% (ES), and the total return rate is 1.15\% for both cases.}
	\tabcolsep=0.18cm
	\begin{tabular}{cccccc}
		\toprule
		& Qinghai& \multirow{2}{*}{GF Securities}& Jingsheng Mechanical & \multirow{2}{*}{Shuanghui}&Arawana \\
        & Salt Lake& & \& Electrical & & Holdings\\
        \midrule
        Sector& Materials& Financials&  Industrials& Consumer Staples&Consumer Staples\\
        Community& 15& 6& 2& 2&1\\
		Weight (VaR)& 0.1309& 0.1500& 0& 0.1500&0.1500  \\
		Weight (ES)& 0.1500& 0.1500& 0.0710& 0.1500&0\\
  	\bottomrule
		\toprule
		& Yunnan & \multirow{2}{*}{NHU}& Guosen & Shenwan &China General   \\
        & Baiyao& & Securities&Hongyuan&Nuclear Power  \\
        \midrule
        Sector& Healthcare& Healthcare& Financials& Financials&Utilities\\
        Community& 10& 1& 1& 8&2 \\
		Weight (VaR)& 0.1500& 0& 0.1500& 0&0.1191 \\
		Weight (ES)& 0& 0.1500& 0.1500& 0.0290&0.1500\\
		\bottomrule
	\end{tabular}
	\label{tab:weightallsec}
\end{table}
\reftab{tab:weightallsec} presents the weight allocations based on VaR and ES as risk measures, along with their corresponding sectors and communities. Stocks with a weight of zero, although included in the maximum independent sets during the optimization process, are not shown in this table. Based on the community labels, we note that three stocks belong to community 1 and another three to community 2. 

Furthermore, we also give an overall portfolio strategy that incorporates the maximum independent sets of all 21 communities. Similar to the sector classification case, we solve the maximum independent set for each community, and aggregate the selected stocks to solve \eqref{eq:portfolio}. This gives the weight allocations in \reftab{tab:weightallcom}. We then compare its performance with that of the sector-based overall portfolio as well as the market portfolio.

\begin{table}[h]
	\centering
	\caption{Optimal portfolio with the minimum risk for the community-based overall stocks; the obtained objective function values are 1.82\% (VaR) and 2.44\% (ES), and the total return rate is 1.15\% for both cases.}
	\tabcolsep=0.25cm
	\begin{tabular}{cccccc}
		\toprule
		& \multirow{2}{*}{Zoomlion}& \multirow{2}{*}{GF Securities}& Gotion& Jingsheng Mechanical &\multirow{2}{*}{Maxwell}\\
        & & & High-tech& \& Electrical &\\
        \midrule
        Community& 1& 6& 2& 2&4\\
		Weight (VaR)& 0.0547& 0.1500& 0.1500& 0 &0 \\
		Weight (ES)& 0& 0.1500& 0& 0.0130 &0.1500 \\
  	\bottomrule
		\toprule
		& Goldwind Science& \multirow{2}{*}{Shuanghui}& \multirow{2}{*}{NHU}& Guosen &China General  \\
        & \& Technology& & &Securities&Nuclear Power \\
        \midrule
        Community& 5& 2& 1& 1&2\\
		Weight (VaR)& 0.1500& 0.1500& 0.0453& 0.1500&0.1500 \\
		Weight (ES)& 0.0870& 0.1500& 0.1500& 0.1500&0.1500 \\
		\bottomrule
	\end{tabular}
	\label{tab:weightallcom}
\end{table}

\begin{figure}[h]
	\centering
	\begin{subfigure}[t]{1\linewidth}
		\centering
		\includegraphics[scale=0.48]{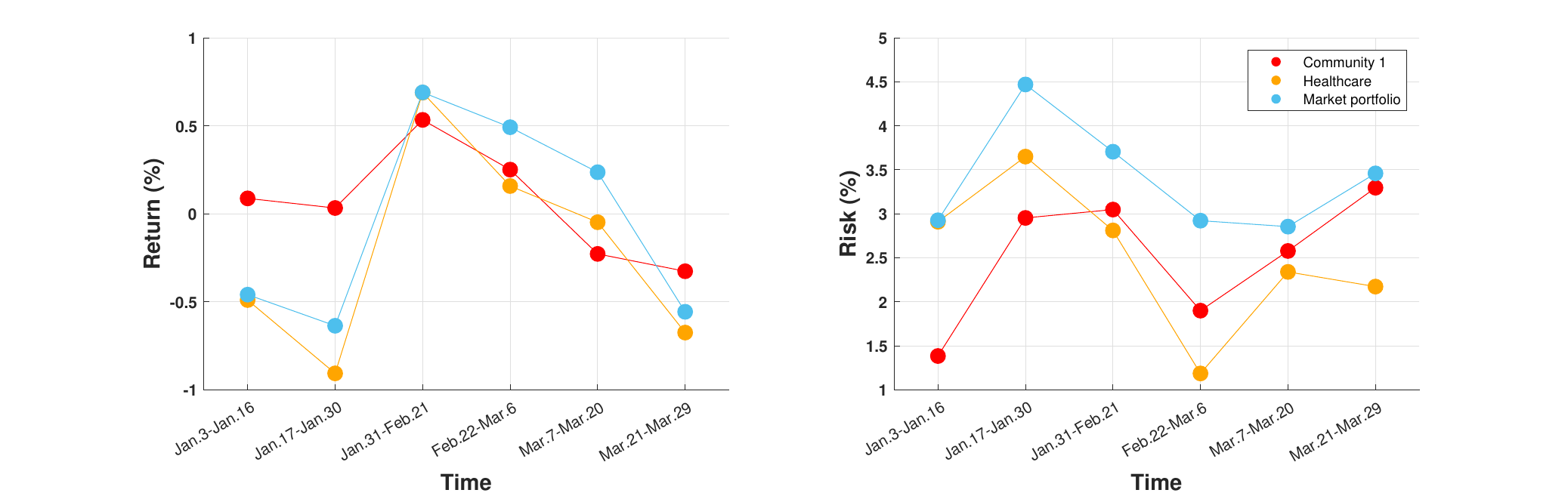}
		\caption{VaR as the risk measure}
	\end{subfigure}
	\begin{subfigure}[b]{1\linewidth}
		\centering
		\includegraphics[scale=0.48]{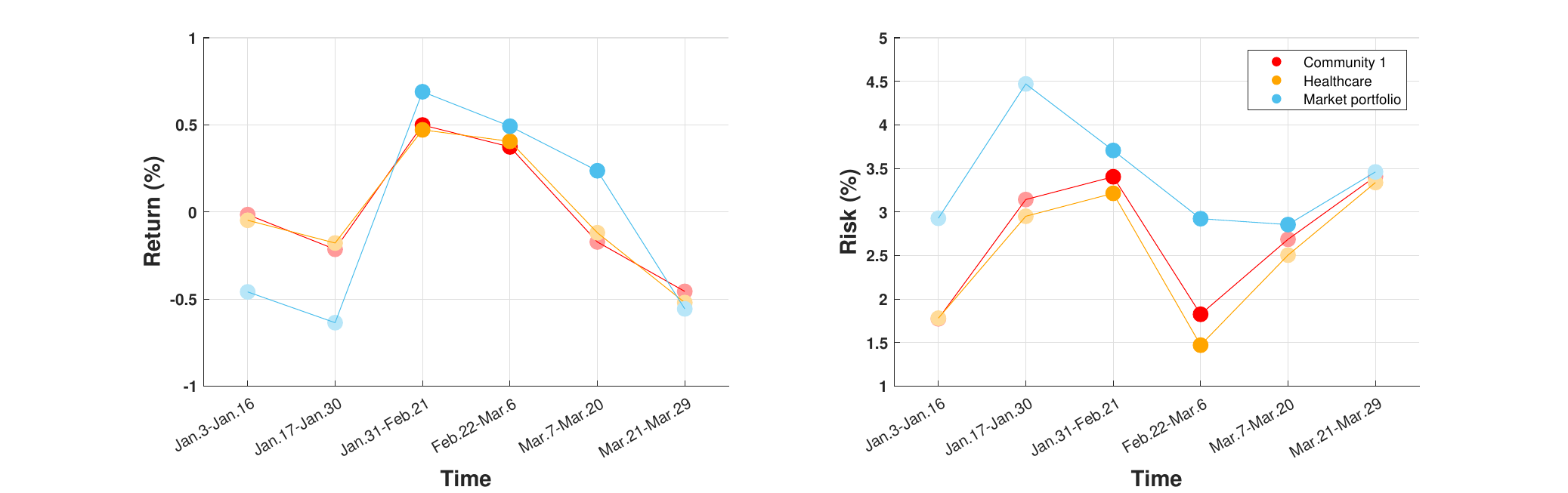}
		\caption{ES as the risk measure}
	\end{subfigure}
	\caption{Line plots of the return-time (left) and risk-time (right) with different risk measure by VaR (a) and ES (b); the red, yellow, and blue nodes respectively represent the community-based overall portfolio, sector-based overall portfolio and the market portfolio.}
	\label{fig:ppall}
\end{figure}

\reffig{fig:ppall} provides comparisons among the performance of community-based and sector-based overall portfolios as well as the market portfolio across different time windows, using both VaR and ES as risk measures. 
In \reffig{fig:ppall}, the community-based overall portfolio (red) demonstrates a significant advantage, especially during a downside market (blue), e.g. Jan 3-16, Jan 17-30, and Mar 21-29, 2024. Moreover, the community-based overall portfolio demonstrates a more stable return profile than the sector-based and the market portfolios, achieving positive returns in several periods (e.g. Jan 3-Mar 6, 2024).

The two right panels of \reffig{fig:ppall} show that the sector-based overall portfolio (yellow) also has a lower risk than the market portfolio. When ES is used as the risk measure, both community- and sector-based strategies give similar performance. However, when using VaR as the risk measure, the left panel of \reffig{fig:ppall}(a) suggests returns of the sector-based portfolio being volatile, which fails to avoid the extremal risk when the market portfolio generates negative returns (e.g. during Jan 3-16 and Jan 17-30, 2024). 
A possible explanation is: that when partitioning the dependence network by sectors, \reffig{fig:dep} shows the existence of excessive isolated nodes so that the corresponding maximum independence set may consist of isolated nodes which actually belong to the same community. In \reftab{tab:weightallsec}, we see that three stocks are classified under community 1 and another three under community 2. If selected stocks belong to the same sector, they are more likely to have edges between them, thereby increasing underlying extremal risk. Such a phenomenon will potentially worsen the portfolio performance in a downside market. In contrast, the community-based overall portfolio effectively mitigates this issue by reducing the likelihood of connections in the maximum independent set in one community. 

For investors who are comfortable with higher risks and seeking potentially greater returns, the local portfolio, with its concentrated investments, may be the better choice. While this portfolio has the potential to deliver a higher maximum return, it also entails increased volatility.
However, for those who prioritize stability and are more risk-averse, the community-based overall portfolio is recommended. This strategy, which involves broad market participation, offers lower risks and more stable performance over time.

\section{Conclusions}\label{sec5}
The core issue addressed in this paper is to provide investors with an investment portfolio that can minimize exposure to extremal risks. We use the extremal dependence measure to quantify the extremal dependence between different stocks and represent the overall dependence structure as a network. Additionally, we use 113 Shenzhen stocks from the CSI 300 as illustrative examples and verify the effectiveness of the proposed strategy.

In our analysis, each stock is regarded as a vertex, and we employ a threshold-based approach to construct the dependence network. To enhance risk diversification, we compare two methods of partitioning the network: sector-based and community-based approaches. We then solve for the maximum independent set within each sector or community and propose a portfolio optimization strategy by minimizing certain risk measures, such as value at risk and expected shortfalls. Furthermore, we evaluate the performance of the portfolios in 2024, comparing both local and overall strategies, and provide investment recommendations tailored to investors' risk tolerance.

For future work, one may consider broadening data selection to gain a better understanding of the extremal dependence structure among stocks, leading to more comprehensive results. It is also essential to consider not only theoretical analysis but also the model's practicality and feasibility in real-world scenarios. This may involve factors such as the impact of investor decisions on market prices, various stock dividend methods, foreign exchange rates, etc. Additionally, the categorization of stocks could be further refined to include common stocks, preferred stocks, and subordinated stocks, among others.

\bibliographystyle{plainnat}
\bibliography{ref.bib}

\end{document}